\documentclass[aps,pra,reprint,nofootinbib,showpacs,superscriptaddress,onecolumn]{revtex4-1}

\usepackage{amssymb}
\usepackage{amsmath, amsthm}
\usepackage{amsfonts}
\usepackage{color}
\usepackage{graphicx}
\usepackage{tikz}
\newtheorem{thm}{Theorem}

\newtheorem{lem}[thm]{Lemma}
\newtheorem{prop}[thm]{Proposition}
\newtheorem{exmp}[thm]{Example}
\newtheorem{rem}[thm]{Remark}
\newcommand{\norm}[1]{\left\Vert#1\right\Vert}
\newcommand{\abs}[1]{\left\vert#1\right\vert}
\newcommand{\set}[1]{\left\{#1\right\}}

\newcommand{\inne}[1]{\left\langle #1\right\rangle}

\newcommand{\complexes}{{\mathbb{C}}}
\newcommand{\C}{{\mathbb{C}}}

\newcommand{\ip}[2]{\langle #1, #2\rangle}

\def\XXint#1#2#3{{\setbox0=\hbox{$#1{#2#3}{\int}$ }
\vcenter{\hbox{$#2#3$ }}\kern-.6125\wd0}}

\newcounter{lastnote}

\begin{document}
	
\title{Equiangular quantum key distribution in more than two dimensions}

\author{Radhakrishnan Balu}
\email{radhakrishnan.balu.civ@mail.mil}
\affiliation{Computer and Information Sciences
	Directorate, Army Research Laboratory, Adelphi, MD 21005-5069.}
\affiliation{Computer Science and Electrical Engineering,
	University of Maryland Baltimore County,
	1000 Hilltop Circle, Baltimore, MD 21250.}
\email{radbalu1@umbc.edu}

\author{Paul J.\ Koprowski}
\email{pkoprowsk@math.umd.edu}
\affiliation{Department of Mathematics \&
	Norbert Wiener Center for Harmonic Analysis and Applications, 
	University of Maryland,
	College Park, MD 20742.}

\author{Kasso A.\ Okoudjou}
\email{kasso@math.umd.edu}
\affiliation{Department of Mathematics \&
	Norbert Wiener Center for Harmonic Analysis and Applications, 
	University of Maryland,
	College Park, MD 20742.}

\author{Justin S.\ Park}
\email{jpark00@mit.edu}
\affiliation{Massachusetts Institute of Technology,
	Cambridge, MA 02139.
}
\author{George Siopsis}
\email{siopsis@tennessee.edu}
\affiliation{Department of Physics and Astronomy, University of Tennessee, Knoxville,
	TN 37996-1200.
}
\date{\today}

\begin{abstract}
We extend the spherical code based key distribution protocols to qudits with dimensions 4 and 16 by constructing equiangular frames and their companions. We provide methods for equiangular frames in arbitrary dimensions for Alice to use and the companion frames, that has one antipode to eliminate one of the possibilities, made up of qudits with $N=4,16$ as part of Bob's code. Non-orthogonal bases that form positive operator valued measures can be constructed using the tools of frames (overcomplete bases of a Hilbert space) and here we apply them to key distribution that are robust due to large size of the bases making it hard for eavesdropping. We demonstrate a method to construct a companion frame for an equiangular tight frame for $\complexes^{p-1}$ generated from the discrete Fourier transform, where $p$ is any odd prime. The security analysis is based on the assumption restricting possible attacks to intercept/resend scenario highlighting the advantages of a qudit over qubit-based protocols.

\end{abstract}

\maketitle

\section{Introduction} \label{intro}
Quantum key distribution (QKD) uses the laws of quantum mechanics to allow
two users to effectively and securely generate a one-time pad in order
to protect sensitive information from adversaries. The first such protocol, the
so-called BB84 algorithm \cite{bennett1984quantum}, employs two sets of
mutually unbiased orthonormal bases of $\C^2$. In this protocol, the first basis set is the eigenbasis of one observable
(for example $\sigma_x$) and the second basis set is the eigenbasis of one of the two sets of complimentary
observables ($\sigma_y$ and $\sigma_z$). In \cite{bruss1998optimal}, the BB84 protocol is extended to six states,
employing both sets of complimentary measurements. The increase in 
the observables allows for better adversarial eavesdropper detection \cite{blow1993fundamental}. Another way to set up quantum communication protocols that minimize error probabilities while maximize mutual information involve non-orthogonal positive-operator valued measures (POVMs) such as the three-state quantum cryptography protocol introduced by Chefles et al \cite{phoenix2000three}. This class of protocols are interesting due to the existence of powerful results on POVMs that can be used to design rejected-data protocols that reveals the presence of eavesdropper using the bits that would be discarded. In \cite{phoenix2000three}, \cite{renes2004spherical} and \cite{renes2005equiangular} the authors move to the 
more general framework of non-orthogonal POVMs for qubit quantum key distribution based on equiangular spherical codes. The simplicity of spherical codes is due to avoidance of sacrificing potential key letters in order to determine the amount of information that an attacker has learned about the key sequence as the success rate provides this information. The protocols further provide a wide range of security and rate of key generation for a given dimension of the systems. When the number of signal states is fixed the spherical codes offer higher noise threshold for security than mutually-unbiased bases with a trade off in terms of lower key generation rates. Specific examples of this family of protocols include two qubit based spherical codes the trine that bests BB84 and the tetrahedron which performs better than six-state in terms of improved resistance to eavesdropping and providing the key error rate in terms of sift rate thus offering a simplified framework. QKD protocols in higher dimensions up to sixteen \cite {Tselniker2009} and the ones based on qudits \cite {Chau2015}, \cite {Cerf2002} that are error-resilient inspired us to look for spherical codes in similar dimensions as they would combine the advantages of both classes of protocols. In this work we are concerned with developing more general spherical codes in higher dimensions using the Hilbertian frames and carry out the security analysis in the context of intercept/resend attacks.

There is a well established correspondence between POVMs and the class of tight frames. Let $d\geq 2.$ A tight frame for $\C^d$  is a set of vectors $\set{f_j}_{j=1}^N \subset \C^d$ such that for all $x\in \C^d$ we
have that $\sum_{j=1}^N \abs{\inne{x, f_j}}^2 = A\norm{x}^2$
for some positive constant $A$. If in addition, $\norm{f_j}=1$,  for each $j=1, \hdots, N$, then $\set{f_j}_{j=1}^N$ is called a \emph{finite unit norm tight frame} (FUNTF), and it is easy to see that $A=N/d$. A FUNTF $F=\set{f_j}_{j=1}^N$  for which there exists a constant $c>0$ with $|\ip{f_j}{f_k}|=c$ for, $j\neq k$ is called an \emph{equiangular tight frame} (ETF) (also known as 
mutual unbiasedness). We refer to \cite{benedetto2003finite, casazza2012finite, OkoudjouFiniteFrame} for more on finite frame theory and some of its applications.
Observe that if $\set{f_j}_{j
=1}^N$ is a FUNTF for $\C^d$, then we can write $$\sum_{j=1}^N \frac{d}
{N} f_j\otimes f^{\dagger}_j = I_{d\times d},$$
which is to say, $\set{\Pi_j=\frac{d}{N}f_j\otimes f^{\dagger}_j}$ forms
a POVM. Similarly, one may construct a unit norm tight frame from any POVM \cite{benedetto2008role}.

Renes' four state protocol \cite{renes2005equiangular} employs a four element ETF  $\set{f_j}_{j=1}^4$
for $\C^2$ with  $\abs{\inne{f_j,f_k}}^2=\frac{1}{3}, ~j\neq k$.
The corresponding POVM is known as a symmetric,
informationally complete, POVM (SIC-POVM). In general, if $N=d^2$ and
$\set{f_j}_{j=1}^{d^2}$ forms an equiangular tight frame for $\complexes^d$, 
then the corresponding POVM is a SIC-POVM. The existence of such ensembles in 
all dimensions is an open problem in harmonic analysis, and quantum information
theory, respectively. Nonetheless, for every dimension $d\geq 2$ there exists an ETF  of $d+1$ vectors in $\C^d$ obtained by taking any $d$ rows of the $(d+1)\times (d+1)$ DFT matrix and renormalizing the resulting column vectors. In the sequel, we shall consider the ETF obtained by taking the last $d$ rows of the  $(d+1)\times (d+1)$ DFT matrix. We call this ETF the $(d+1, d)$ Fourier ETF, or simply the Fourier ETF when the context is clear. More generally, using a difference set sampling strategy, the class of harmonic equiangular tight frames may be constructed (cf. \cite{xia2005achieving}).

Both the three state and four state quantum key algorithms rely on a 
measurement ensemble, generated by a companion equiangular tight frame $\set{g_j}$ defined
as follows: given an equiangular tight frame $F=\set{f_j}_{j=1}^N$, the equiangular
tight frame $G=\set{g_j}_{j=1}^N$ is a {\it companion equiangular tight frame} for $F$ if
\begin{equation} \label{ek}
	\abs{\inne{g_j, f_k}}^2 = \begin{cases}
			0 & k=j \\
			c & \textrm{otherwise}
		\end{cases}
\end{equation}
Much like the existence of equiangular frames, the construction of such
sets is a non-trivial problem. In this paper we offer  constructions of  companion equiangular tight frames to the $(d+1, d)$ Fourier ETF for a family of values of $d$. We then extend the equiangular QKD algorithms to these dimensions, and illustrate our algorithms with some examples.

For completeness, we recall the set up of the equiangular QKD  protocol. Assume that  Alice and Bob wish to communicate securely and have access to a
quantum channel as well as a classical one.
Alice and Bob predetermine an equiangular frame set of states $\set{f_j}_{j=1}^N$
from which Alice uniformly samples from the $N$ states and picks out
$f_k$, which she sends to Bob.
Bob has a measurement device corresponding to the POVM $\set{G_j= 
\frac{d}{N} g_j\otimes g^\dagger_j}_{j=1}^N$ where $\set{g_j}_{j=1}^N$ is a 
companion equiangular frame for $\set{f_j}_{j=1}^N$.
Bob receives $f_k$ from Alice and performs a measurement with outcome $l\in\set{1,...,N}$.
Now Bob knows with certainty, Alice did not send $f_l$, as the 
probability of measuring $l$ given $f_l$ is $\abs{\inne{g_l,f_l}}^2
=0$. However, Bob knows nothing about which of
the other $N-2$ possible states that might have been sent.
To determine this, Bob then communicate a 
random sampling $S$ of $N-2$ elements of $\set{1,...,N}\setminus\set
{l}$ without replacement. He sends the sample $S$ to Alice
through a classical channel. If $k\in S$, then Alice signals failure and sends a new quantum
state. If $k\not\in S$ (which has a probability of $\frac{1}{N-1}$ of happening) then Alice and Bob
both know that Alice sent state $k$, while anyone viewing the classical communication only
knows that Alice sent either $f_k$ or $f_l$. Alice and Bob generate a random classical bit based
on an a priori agreed upon algorithm (say $b=1$ if $(-1)^l=1$ and $b=0$ otherwise).
Based on eavesdropping of the classical channel, an eavesdropper Eve has at best a $2^{-k}$
probability of guessing the correct $k$ bit number based on complete knowledge of the classical
communications, which would presumably have some sort of classical encryption. Similarly, an intercept and resend attack  on the quantum channel would quickly be detected, as Alice and Bob's keys would not match with arbitrarily high probability.

Before the difficulty of experimental implementation, there is the 
non-trivial task of generating equiangular frames, and the associated companion set. In $\C^2$, the geometric representation of the Bloch sphere was used in order to construct such sets \cite{phoenix2000three, renes2004spherical}. However, this type of  geometric construction seems absent in higher dimensions. Nonetheless we shall construct a family of companion ETFs starting from some $(d+1, d)$ Fourier ETFs.

We demonstrate later that, when $d+1$ is any odd prime, a $(d+1, d)$ Fourier ETF $F=\set{f_j}_{j=1}^{d+1}$ for $\C^{d}$ and a $d \times d$ diagonal unitary and traceless matrix $U$ exist such that
\begin{equation*}
G = \set{g_j| g_j=Uf_j, j=1,...,d+1}
\end{equation*}
is a companion equiangular frame for $F$.

This is easily accomplished in two dimensions using the Bloch sphere
representation and doing a three dimensional rotation 
within that representation and mapping back to $\C^2$. For example, let $f_j=\frac{1}{\sqrt{2}}[1~ e^{i\pi j/3}]^*$ for $j=0,1,2$. Then the transformation
$$R=\begin{bmatrix}
     1 & 0\\ 0 & e^{i\pi}
    \end{bmatrix},$$
which amounts to a 180 degree rotation in the $xy$ plane in the Bloch sphere, accomplishes the desired
result: \begin{align}\abs{\inne{Rf_j, f_k}}^2=\begin{cases}
                                  0 & j=k \\ \frac{3}{4} & j\neq k
                                 \end{cases}.\label{ek2}\end{align}

If $F=\{f_j\}_{j=1}^N$ is an ETF for $\C^d$ and if there exists a companion ETF $G=\{g_j=Uf_j\}_{j=1}^N$ for some unitary $d\times d$ matrix $U$, then we may proceed
in generalizing Renes' protocol. In particular, the common inner product of $F$ (hence of $G$) is $\alpha=\frac{N-d}{d(N-1)}$. The frame operators of $F$ and $G$ are also identical, and equal  $N/d I_{d\times d}$. Hence we may
define a positive-operator valued measure (POVM) associated with each frame as $G_j=\frac{d}{N}g_jg_j^*$ and $F_j=\frac{d}{N}f_jf_j^*.$

Suppose Alice prepares a state $f_k$ and sends it to Bob. If Bob then measures using the $G_j's$ then the probability
of measuring outcome $j$ in an experiment is given by
\begin{equation}\label{ek4}
Pr(j|f_k)  = tr(G_j f_kf_k^*) = tr(f_k^*G_j f_k)= \frac{d}{N} \inne{g_j,f_k}\inne{f_k,g_j}=\frac{d}{N}\abs{\inne{g_j,f_k}}^2.
\end{equation}

Now, using the fact that the $f_k's$ form an $N/d$ tight frame, that $g_j$ has a unit norm, and
that the sets satisfy Equation \eqref{ek} we have for $j\neq k$
\begin{equation*}
\abs{\inne{g_j,f_k}}^2 =\frac{1}{N-1} \sum_{k\neq j}\abs{\inne{g_j,f_k}}^2  =\frac{N}{d(N-1)}\norm{g_j}^2  =\frac{N}{d(N-1)}.
\end{equation*}
Combining with Equation (\ref{ek4}) yields
$$pr(j|f_k)=\begin{cases} 0 & j=k\\ \frac{1}{N-1} & j\neq k\end{cases}.$$
Hence, for a fixed measurement outcome $j$, there is an equal probability
that the state being measured was $f_k$ for $k\neq j$ and no probability that
the state was $f_j$.

In some case, there might not exist a unitary matrix $U$ that would produce a companion ETF $G=\{Uf_j\}_{j=1}^N$ from  an ETF $F=\{f_j\}_{j=1}^N$ for $\C^d$. Indeed, Renes also has a four element equiangular frame given by
$$F=\begin{bmatrix}
   \alpha & \alpha & \beta & \beta \\
   i\beta & -i\beta & \alpha & \alpha
  \end{bmatrix}$$
where $\alpha = \sqrt{\frac{1}{6}(3+\sqrt{3})}$ and $\beta = \sqrt{\frac{1}{6}(3-\sqrt{3})}.$
Let $$U=\begin{bmatrix}
       a & b\\ c & d
      \end{bmatrix}$$ be unitary. Then solving $diag(F^*UF)=[0, 0, 0, 0]$ non-trivially is actually impossible
      as it requires $b=c$ which implies $a=d=0$ which implies $b=-c$ or similar contradiction. Therefore,
no unitary $U$ exists such that $g_j=Uf_j$ exists that satisfies~\eqref{ek}.
However, if we set $a=d=0$ and $b=c=1$ then $g_j=Uf_j$ for $j=1,2$ and $g_3=Uf_4$ and $g_4=Uf_3$
then $g_j$ and $f_j$ satisfy~\eqref{ek}. One can ask whether such a unitary transformation (up to re-indexing) exists for higher dimensions.
If it does, then we can generalize the two dimensional results from Renes to arbitrary
higher finite dimensions. Namely, if such an $R$ works in dimension $d$, we would have
$g_j= Rf_j$ in~\eqref{ek} and our measurement operators would be scaled versions of $g_jg_j^*$. Therefore, a companion ETF can be constructed if one can find  a unitary transformation
$U$ and a permutation matrix $P$ such that $G=UFP$ where $F$ is the matrix synthesis operator of the initial frame
and $G=[g_1, g_2,..., g_{N}]$ is the synthesis operator for the desired new frame.
Hence,~\eqref{ek} may be reformulated as $$\abs{(G^*PFU)_{i,j}}^2=\abs{(P^*F^*U^*G)_{i,j}}^2=\begin{cases}
0 & i=j\\ c & o.w.
\end{cases}.$$ 


The main goal of this paper is to construct companion ETF from the $(d+1, d)$ Fourier ETF when $d+1$ is prime. This is achieved by constructing a  $d\times d$ traceless diagonal matrix of $\pm 1$. Let $\tilde{u}\in \C^d$ be the vector of $\pm 1$ consisting of the diagonal entries of $U$, and $u=\begin{bmatrix}0\\ \tilde{u}\end{bmatrix} \in \C^{d+1}$. Then $u$ is an eigenvector of $W$, the $(d+1)\times (d+1)$ DFT matrix. We conjecture that every unitary diagonal traceless matrix $U$ yielding a companion ETF to the $(d+1, d)$ Fourier ETF  necessarily generates either an eigenvector of the DFT matrix as described above, or a vector $u$ such that $Wu=\lambda u^*$ for sume unimodular number $\lambda$. We have not been able to prove this conjecture, but through exhaustive search, we observed that there indeed exist such vector for all prime number up to $59$. Furthermore, our numerical search shows that no such eigenvector exist for  composite numbers  in this range.

\section{Companion ETF in prime dimensions}\label{sec:sec2}



As mentioned in the introduction, starting from the ETF $\{f_k\}_{k=0}^2\subset \C^2$, it is known that the family $\{Rf_k\}_{k=0}^2$ is a companion ETF where $R=\begin{pmatrix}1&0\\0&-1\end{pmatrix}$. Note that $R$ is unitary and traceless. One is naturally lead to ask if, given a FUNTF $\{f_k\}_{k=1}^{N}\subset \C^d$, can one find a unitary traceless $d\times d$ matrix $U$ such that $\{f_k\}_{k=1}^{N}$ and $\{Uf_k\}_{k=1}^{N}$ are companion ETFs. 

Before we answer this question in some special cases, we note that if $\set{f_j}_{j=1}^N$ is an equiangular FUNTF for $\C^d$,
the set of $N^2$ $d\times d$ matrices defined by $\set{f_j\otimes f_k=f_jf_k^*}_{j,k=1}^N $ forms a two distance tight frame for $C^{d\times d}$ under the Hilbert Schmidt inner product, \cite{BGOY15}. 

\begin{prop}\label{prop1}
Suppose that $\set{f_j}_{j=1}^N$ is an equiangular FUNTF for $\C^d$. Then $$\{f_j\otimes f_k\}_{j, k=1}^N=\{f_jf_k^*\}_{j, k=1}^N$$ is a two-distance FUNTF for $ \C^{d\times d}$ under the Hilbert Schmidt
inner product.

\end{prop}

\begin{proof}

We have from the properties of the tensor product that
$$\inne{f_m\otimes f_n, f_j\otimes f_l}_{HS}= \inne{f_m f_n^*, f_jf_l^*}_{HS}=\inne{f_m, f_j}\inne{f_l,f_n}.$$
Since $\abs{\inne{f_j, f_k}}^2=\alpha=\frac{N-d}{d(N-1)}$ for all $j\neq k$, we have that
$$\abs{\inne{f_m\otimes f_n, f_j\otimes f_l}_{HS}}^2=\begin{cases}
1 & m=j, ~l=n\\ \alpha & m=j,~l\neq n\\ \alpha & m\neq j, ~ l=n\\ \alpha^2 & m\neq j, ~ l\neq n
\end{cases}.$$ So we have a unit normed two distance (in absolute value) set (distances $\alpha$ and $\alpha^2$) with 
$\alpha$ occurring $2(N^2)(N-1)$ times, $\alpha$ occurring $N^2(N-1)^2$ times. To show tightness,
let $M\in \C^{d\times d}$ be arbitrary. We have
\begin{equation}\label{ek3}
 \sum_j\sum_k \abs{\inne{M,f_jf_k^*}_{HS}}^2  = \sum_j\sum_k \abs{tr(f_j^*Mf_k)}^2\ = \sum_k\sum_j \abs{\inne{Mf_k, f_j}}^2 = \frac{N}{d}\sum_k\norm{Mf_k}^2.
\end{equation}
We also have for $M^*=[M_1,...,M_d]$ that $\norm{Mf_k}^2=\sum_\ell \abs{\inne{f_k, M_\ell}}^2$ and
therefore
\begin{align*}
 \sum_k\norm{Mf_k}^2 & = \sum_\ell\sum_k \abs{\inne{f_k, M_\ell}}^2 =\sum_\ell \frac{N}{d} \norm{M_\ell}^2  \\ 
 &=\frac{N}d \sum_{l,i}\abs{M_l[i]}^2  =\frac{N}d \norm{M}_{HS}^2.
\end{align*}
Plugging into  \eqref{ek3} shows that
$\set{f_k\otimes f_j}_{j,k\in\set{1,...,N}}$ is a $N^2/d^2$ tight frame for $\C^{d\times d}$.
\end{proof}

Proposition~\ref{prop1} can be used as follows. 
If $F=\{f_j\}_{j=1}^N$ is an ETF for $\C^d$, then to find a unitary $d\times d$ matrix $U$ such that $G=\{g_j=Uf_j\}_{j=1}^N$ is a companion ETF to $F$ reduces to finding the coefficients $(\ip{U}{f_j\otimes f_\ell}_{HS})_{j, \ell=1}^N$. However, $$\ip{U}{f_j\otimes f_\ell}_{HS}=\textrm{tr}(Uf_\ell f_j^*)=\textrm{tr}(f_j^*Uf_\ell)=\ip{Uf_\ell}{f_j}=\ip{g_\ell}{f_j}=\sqrt{\alpha}\, e^{2\pi i \theta_{\ell, j}}$$ where $\alpha=\frac{N-d}{d(N-1)}$, and $\theta_{\ell, j}\in [0, 1)$ is an unknown phase factor. Thus, determining $U$ is equivalent to finding these unknown phases. This is an example of the nontrivial phase retrevial problem, see \cite{Bal16} and the references therein for more details. From a complexity point of view, $U$ belongs to the $d^2$ dimensional space $\C^{d\times d}$ for which $\{f_j\otimes f_k\}_{j, k=1}^N=\{f_jf_k^*\}_{j, k=1}^N$ is a two-distance FUNTF of $N^2$ vectors. The right regime to recover $U$ from only the magnitudes of its frame coefficients is $N^2>d^4$, i.e., $N>d^2$. But as we shall see, the results we obtain are for $N=d+1$. Consequently, our results are not covered by the phaseless reconstruction theory. 

Because of the complexity of the problem, we seek a unitary, diagonal and traceless $d\times d$ matrix that would produce a companion ETF from an ETF $F$. In particular, we shall only consider the case where $F$ is the $(d+1, d)$ Fourier ETF, and show that finding such diagonal unitary matrix reduces to finding a specific eigenvector of the DFT matrix.

\subsection{Construction of companion FUNTFs in prime dimensions}\label{subsec2.2}
Let $d\geq 2$ be fixed and set $\omega = e^{-\frac{2\pi i}{d}}$. Suppose that  $F=\set{f_k}_{k=1}^{d+1}$ is a $(d+1, d)$ Fourier ETF   for $\C^{d}$ generated by
taking the columns of the $(d+1)$-dimensional DFT matrix, removing the top row and scaling by $\frac{1}{\sqrt{d}}$.
Let $v[k]$ denote the $k$-th entry in the vector $v$, starting with 0 (so $v[0]$ is the leading entry).
Assume there exists a traceless, diagonal, unitary $d \times d$ matrix $U$ such that
$\abs{\inne{Uf_k,f_j}} = \begin{cases} 0 & k=j \\ c & o.w. \end{cases}$.
We recall that $c = \frac{\sqrt{d+1}}{d}$,
and we have for $k\neq j$ that
\begin{align} 
\abs{\inne{Uf_k,f_j}} &= \abs{\sum_{n=1}^{d} (Uf_k)[n-1] \overline{f_j[n-1]}} = \abs{\sum_{n=1}^{d} U_{n,n}f_k[n-1] \overline{f_j[n-1]}} \notag  \\
&= \frac{1}{d}\abs{\sum_{n=1}^{d} U_{n,n} \omega^{n(k-j)}} = \frac{1}{d}\abs{\sum_{n=1}^{d} U_{n,n} \omega^{n\ell}} \notag \\
&= \frac{1}{d}\abs{\sqrt{d+1}},
 \label{ft}
\end{align}
where $\ell=k-j\neq 0$. Hence, if we denote the diagonal $D$ of $U$ as
$D = \begin{bmatrix} U_{1,1} \\ \vdots \\ U_{d,d}\end{bmatrix}$
and embed $D$ in $\C^{d+1}$ via the mapping
\begin{equation*}
D \mapsto \begin{bmatrix} 0 \\ D \end{bmatrix} = f,
\end{equation*}
then \eqref{ft} implies that for $\ell\neq 0$
\begin{equation} \label{fixedR}
\abs{\widehat{f}[\ell]} = \abs{\frac{1}{\sqrt{d+1}}\sum_{n=0}^{d} \omega^{n\ell} f[n]} = \frac{1}{\sqrt{d+1}}\abs{\sum_{n=1}^{d} U_{n,n} \omega^{n \ell}}= 1,
\end{equation}
where $\widehat{f}$ is the ($d+1$)  DFT of $f$.
Since $U$ is unitary, we have that $\abs{f[j]} = 1$ for $j\neq 0$. Furthermore, the traceless condition on $U$ implies that $\hat{f}[0]=f[0]=0$. Thus the vector $f$ and its DFT $\hat{f}$ have unimodular entries except their first entry which is $0$. Because the eigenvalues of the ($d+1\times d+1$) DFT matrix are $\pm 1, \pm i$, it is clear that the corresponding eigenvectors $u$ have the property that $|\hat{u}[k]|=|u[k]|$ for $k=0, \hdots, d$. Therefore,  if we find a function of the form of $f$ that is an eigenfunction of the ($(d+1)\times (d+1)$) DFT, then the lower $d$ unit modulus entries of $f$ define
a traceless, diagonal, unitary transformation that generates a companion equiangular tight frame for $\set{f_k}_{k=1}^{d+1}$. The following  construction of such an eigenvector is given in \cite{horn2010interesting}, when $d+1=p$ is a prime odd number. In the sequel we denote the $p\times p$ DFT matrix by $W$. We refer to \cite{horn2010interesting} for a proof. 

\begin{prop}\label{egenvecdft}
Define $f\in \C^p$ by 
\begin{equation*}
f = \left[ 0, \left(\frac{1}{p}\right)_2, \left(\frac{2}{p}\right)_2, \cdots, \left(\frac{k}{p}\right)_2, \cdots, \left(\frac{p-1}{p}\right)_2 \right]^*
\end{equation*}
where $\left(\dfrac{n}{p}\right)_2$ is the Legendre symbol, defined by
\begin{equation*}
\left(\frac{n}{p}\right)_2 = \begin{cases}
1 & \textrm{if $n$ is a quadratic residue modulo $p$} \\
-1 & \textrm{if $n$ is not a quadratic residue modulo $p$}
\end{cases}
\end{equation*}
for $1 \leq n \leq p-1$.
Then $f$ is an eigenvector of $W$. Furthermore, when $p \equiv 1$ (mod 4), the eigenvalue for this vector is 1, and when $p \equiv 3$ (mod 4), the eigenvalue is $-i$.
\end{prop}


In fact, our main result shows that this is the only eigenvector of the form $[0, \pm 1, \pm 1, \cdots, \pm 1]$ for $W$.  More specifically,

\begin{thm}\label{mainthm}
	If $u_1, u_2$ are eigenvectors of $W$ of the form $[0, 1, \pm 1, \cdots, \pm 1]^*$, then $u_1 = u_2$.
\end{thm}

The proof of this result is based on the following lemmas, which we first prove. For simplicity and without loss of generality, the following proofs standardize the vectors by assuming that the first nonzero entry is +1.

\begin{lem} \label{lemma1}
	If $u_1, u_2$ are distinct vectors of the form $[0, 1, \pm 1, \cdots, \pm 1]^*$ such that $Wu_1 = \lambda_1 u_1$ and $Wu_2 = \lambda_2 u_2$, then $\lambda_1 \neq \pm \lambda_2$.
\end{lem}

\begin{proof}
	Assume for the sake of contradiction that $\lambda_1 = \lambda_2$. (The $\lambda_1 = -\lambda_2$ case is shown similarly.)
	
	From the first row of $W$,
	\begin{equation*}
	u_1[1] + u_1[2] + u_1[3] + \cdots + u_1[p-1] = 0
	\end{equation*}
	and
	\begin{equation*}
	u_2[1] + u_2[2] + u_2[3] + \cdots + u_2[p-1] = 0
	\end{equation*}
	Define $v[k] = (u_1[k] - u_2[k])/2$ for $1 \leq k \leq p-1$. Then by subtracting the second equation from the first and dividing by 2,
	\begin{equation} \label{eq:vSum}
	v[1] + v[2] + v[3] + \cdots + v[p-1] = 0.
	\end{equation}
	From the second row of $W$,
	\begin{equation*}
	u_1[1]\omega + u_1[2]\omega^2 + u_1[3]\omega^3 + \cdots + u_1[p-1]\omega^{p-1} = u_1[1]\lambda_1 = \lambda_1
	\end{equation*}
	and
	\begin{equation*}
	u_2[1]\omega + u_2[2]\omega^2 + u_2[3]\omega^3 + \cdots + u_2[p-1]\omega^{p-1} = u_2[1]\lambda_2 = \lambda_2 = \lambda_1.
	\end{equation*}
	By subtracting the second equation from the first and dividing by 2,
	\begin{equation} \label{eq:vRootSum}
	v[1]\omega + v[2]\omega^2 + v[3]\omega^3 + \cdots + v[p-1]\omega^{p-1} = 0.
	\end{equation}
	Let $A = \{ k : v[k] = 1, 1 \leq k \leq p-1 \}$, $B = \{ k : v[k] = -1, 1 \leq k \leq p-1 \}$, and $C = \{ 0, 1, 2, \cdots, p-1 \} \setminus B$. By a basic property of roots of unity,
	\begin{equation*}
	\sum_{k \in B} \omega^k + \sum_{k \in C} \omega^k = 0.
	\end{equation*}
 \eqref{eq:vRootSum} can be written as
	\begin{equation*}
	\sum_{k \in A} \omega^k - \sum_{k \in B} \omega^k = 0.
	\end{equation*}
	Combining the two equations above,
	\begin{equation} \label{eq:rootSum}
	\sum_{k \in A} \omega^k + \sum_{k \in C} \omega^k = 0.
	\end{equation}
	 \eqref{eq:vSum} implies that $|A| = |B|$. Then $|A| + |C| = |A| + p - |B| = |A| + p - |A| = p$.
	Note that $A$ and $B$ are disjoint, so $A$ and $C$ are not. Thus, \eqref{eq:rootSum} is a vanishing \textit{asymmetric} sum of $p$ $p$-th roots of unity.
	However, this is not possible by  \cite[Theorem 3.3]{lam2000vanishing}, raising a contradiction. Therefore, $\lambda_1 \neq \lambda_2$.
\end{proof}

\begin{lem} \label{lemma2}
	If $u_1, u_2$ are distinct vectors of the form $[0, 1, \pm 1, \cdots, \pm 1]^*$ such that $Wu_1 = \lambda_1 u_1$ and $Wu_2 = \lambda_2 u_2$, then $\lambda_1 \neq \pm i\lambda_2$.
\end{lem}

\begin{proof}
	Assume for the sake of contradiction that $\lambda_1 = i\lambda_2$. (The $\lambda_1 = -i\lambda_2$ case is shown similarly.)
	
	From the second row of $W$,
	\begin{equation*}
	u_1[1]\omega + u_1[2]\omega^2 + u_1[3]\omega^3 + \cdots + u_1[p-1]\omega^{p-1} = u_1[1]\lambda_1 = \lambda_1
	\end{equation*}
	and
	\begin{equation*}
	u_2[1]\omega + u_2[2]\omega^2 + u_2[3]\omega^3 + \cdots + u_2[p-1]\omega^{p-1} = u_2[1]\lambda_2 = \lambda_2 = -i\lambda_1.
	\end{equation*}
	Let $A_1 = \{ k : u_1[k] = 1, 1 \leq k \leq p-1 \}$, $B_1 = \{ k : u_1[k] = -1, 1 \leq k \leq p-1 \}$, and $C_1 = \{ 0, 1, 2, \cdots, p-1 \} \setminus B$. Then $|A_1| + |C_1| = p$, and by following the process in Lemma \ref{lemma1},
	\begin{equation*}
	\sum_{k \in A_1} \omega^k + \sum_{k \in C_1} \omega^k = \lambda_1.
	\end{equation*}
	Similarly, by letting $A_2 = \{ k : u_2[k] = 1, 1 \leq k \leq p-1 \}$, $B_2 = \{ k : u_2[k] = -1, 1 \leq k \leq p-1 \}$, and $C_2 = \{ 0, 1, 2, \cdots, p-1 \} \setminus B$, it follows that $|A_2| + |C_2| = p$ and
	\begin{equation*}
	\sum_{k \in A_2} \omega^k + \sum_{k \in C_2} \omega^k = -i\lambda_1.
	\end{equation*}
	Let $\omega_0 = e^{\frac{-\pi i}{2p}}$, so $\omega_0$ is a $4p$-th root of unity such that $\omega_0^4 = \omega$. Then the previous two equations are equivalent to
	\begin{equation} \label{eq:rootSum1}
	\sum_{k \in A_1} \omega_0^{4k} + \sum_{k \in C_1} \omega_0^{4k} = \lambda_1
	\end{equation}
	and
	\begin{equation*}
	\sum_{k \in A_2} \omega_0^{4k} + \sum_{k \in C_2} \omega_0^{4k} = -i\lambda_1,
	\end{equation*}
	respectively. Multiplying the second equation by $-i = e^{\frac{-\pi i}{2}} = \omega_0^p$,
	\begin{equation} \label{eq:rootSum2}
	\sum_{k \in A_2} \omega_0^{4k+p} + \sum_{k \in C_2} \omega_0^{4k+p} = -\lambda_1.
	\end{equation}
	Adding  \eqref{eq:rootSum1} and \eqref{eq:rootSum2},
	\begin{equation} \label{eq:fullRootSum}
	\sum_{k \in A_1} \omega_0^{4k} + \sum_{k \in C_1} \omega_0^{4k} + \sum_{k \in A_2} \omega_0^{4k+p} + \sum_{k \in C_2} \omega_0^{4k+p} = 0.
	\end{equation}
	This is a sum of $2p$ $4p$-th roots of unity. Since $p$ is an odd prime, it follows from \cite[Theorem 3.3]{lam2000vanishing} that such a sum must be
	one of:
	\begin{itemize}
		\item $p$ symmetric sums of two $4p$-th roots of unity, or
		\item two symmetric sums of $p$ $4p$-th roots of unity.
	\end{itemize}
	We now show that both of these are impossible.
	
	Choose any $k$ in $A_1 \cup C_1$. Since $p$ is odd, $k+\frac{p}{2}$ cannot be in $A_1 \cup C_1$ and $k+\frac{p}{4}$ cannot be in $A_2 \cup C_2$, so $\omega^{4k}$ is in the sum but $-\omega^{4k} = \omega^{4k+2p}$ is not.
	Thus, the sum cannot consist of $p$ symmetric sums of two $4p$-th roots of unity.
	
	Since $|A_1| + |C_1| = p$ and $A_1$ and $C_1$ are not disjoint, the sum in Equation \ref{eq:rootSum1} is not a symmetric sum of $p$ roots of unity.
	However, every term in this sum is a $p$-th root of unity, while no term in \eqref{eq:rootSum2} is a $p$-th root of unity.
	Thus, the sum in \eqref{eq:fullRootSum} cannot consist of two symmetric sums of $p$ $4p$-th roots of unity. 
	
	The sum in  \eqref{eq:fullRootSum} is neither $p$ symmetric sums of two $4p$-th roots of unity nor two symmetric sums of $p$ $4p$-th roots of unity,
	which gives the desired contradiction. Therefore, $\lambda_1 \neq i\lambda_2$.
\end{proof}

We are now ready to prove Theorem~\ref{mainthm}. 

\begin{proof}{Proof of Theorem~\ref{mainthm}.}
	Let $Wu_1 = \lambda_1 u_1$ and $Wu_2 = \lambda_2 u_2$. Since the only eigenvalues of the DFT are $1, -1, i, \textrm{ and } -i$, either $\lambda_1 = \pm \lambda_2$ or $\lambda_1 = \pm i\lambda_2$.
	If $u_1 \neq u_2$, then these are both impossible according to Lemmas \ref{lemma1} and \ref{lemma2}. Therefore, $u_1 = u_2$.
\end{proof}

Using this construction, an equiangular tight frame $F = \set{f_j}_{j=1}^N$ for $\C^{N-1}$ along with a companion frame $G$ can be constructed for any prime $N=p+1$.
In particular, the companion frame satisfies
\begin{equation*}
    G = \set{g_j| g_j=Uf_j, j=1,...,N}
\end{equation*}
where $U$ is the $(N-1) \times (N-1)$ matrix whose diagonal entries are the lower $N-1$ entries in $f$.

By an exhaustive computational search, the existence and uniqueness of the eigenvector in the above construction was verified for all primes up to 59.
Interestingly, the search yielded no eigenvectors of the form $[0, \pm 1, \pm 1, \cdots, \pm 1]^*$ for composite $N$ up to this same value, and we conjecture that no such eigenvector exists for any composite $N$.
While this fact is evident if $N$ is even (one need simply consider the first row of the DFT), a full proof of this fact is not forthcoming.

\begin{exmp}\label{example1}
We provide a few examples of the construction above.
We construct an equiangular tight frame $\set{f_j}_{j=1}^4$ in $\complexes^4$ by sampling the $5\times 5$ DFT matrix.
Indeed, we have
\begin{equation*}
    DFT =\tfrac{1}{\sqrt{5}} \begin{bmatrix}
                1 & 1 & 1 & 1 & 1\\
                1 & \omega\phantom{1} & \omega^2 & \omega^3 & \omega^4\\
                1 & \omega^2 & \omega^4 & \omega\phantom{1} & \omega^3\\
                1 & \omega^3 & \omega\phantom{1} & \omega^4 & \omega^2\\
                1 & \omega^4 & \omega^3 & \omega^2 & \omega\phantom{1}
            \end{bmatrix},
\end{equation*}
and
\begin{equation*}
    P = \begin{bmatrix}
            0 & 1 & 0 & 0 & 0\\
            0 & 0 & 1 & 0 & 0\\
            0 & 0 & 0 & 1 & 0\\
            0 & 0 & 0 & 0 & 1
        \end{bmatrix},
\end{equation*}
and we set $f_j$ equal to the $j^{th}$ column of $\frac{1}{2}P*DFT$.
Define $g_j=Uf_j$ for $j=1,...,5$ where
\begin{equation*}
    U = \begin{bmatrix}
            1 & 0 & 0 & 0\\
            0 & -1 & 0 & 0 \\
            0 & 0 & -1 & 0 \\
            0 & 0 & 0 & 1
        \end{bmatrix}.
\end{equation*}
Then we have $\inne{g_j,f_j} = 1-1-1+1=0$ and $\abs{\inne{g_k, f_l}}^2=\frac{5}{16}$ for $k \neq l$.
Hence, $\set{g_j}_{j=1}^4$ is a companion equiangular frame for $\set{f_j}_{j=1}^4$.

Similarly, sampling the $7 \times 7$ DFT matrix and employing $U = diag[1,1,-1,1,-1,-1]$
generates an equiangular harmonic frame and a companion equiangular frame for $\complexes^{6}$, where
$\inne{g_j,f_j} = 1+1-1+1-1-1=0$ and $\abs{\inne{g_k, f_l}}^2=\frac{7}{36}$ for $k \neq l$.
\end{exmp}
%

\begin{rem}
When $p \equiv 1 \mod 4$ is prime, \cite{horn2010interesting} provides a second construction which satisfies the criteria for $f$.
The vector is
\begin{equation*}
f = \left[ 0, \left(\frac{1}{p}\right)_4, \left(\frac{2}{p}\right)_4, \cdots, \left(\frac{k}{p}\right)_4, \cdots, \left(\frac{p-1}{p}\right)_4 \right]^*
\end{equation*}
where $\left(\dfrac{n}{p}\right)_4$ is defined by
\begin{equation*}
\left(\frac{n}{p}\right)_4 = \begin{cases}
    1 & \textrm{if $n^{(p-1)/4} \equiv 1 \mod p$} \\
    i & \textrm{if $n^{(p-1)/4} \equiv c \mod p$} \\
    -1 & \textrm{if $n^{(p-1)/4} \equiv c^2 \equiv -1 \mod p$} \\
    -i & \textrm{if $n^{(p-1)/4} \equiv c^3 \equiv -c \mod p$}
\end{cases}
\end{equation*}
Here, $c$ is defined as a primitive fourth root of unity in the multiplicative group of integers mod $p$, i.e. an integer $c$ such that $c^2 \equiv -1 \mod p$.

While this $f$ is not an eigenvector of the DFT, it still satisfies the property that each entry except for the first has magnitude 1
and that the magnitude of each entry remains fixed under the DFT.
In particular, there exists a complex constant $z$ of magnitude 1 such that $Wf = z\overline{f}$.
Thus, as in the previous construction, this vector $f$ can be used to construct a diagonal matrix $U$ which generates a companion frame.

As an example of this construction, sampling the $5 \times 5$ DFT matrix and employing $U = diag[1,i,-i,-1]$
generates an equiangular harmonic frame and a companion equiangular frame for $\complexes^{4}$, where
$\inne{g_j,f_j} = 1+i-i-1=0$ and $\abs{\inne{g_k, f_l}}^2=\frac{5}{16}$ for $k \neq l$.

In fact, as a generalization of the above constructions, if p is a prime number congruent to 1 mod m, then define the vector
\begin{equation*}
f = \left[ 0, \left(\frac{1}{p}\right)_m, \left(\frac{2}{p}\right)_m, \cdots, \left(\frac{k}{p}\right)_m, \cdots, \left(\frac{p-1}{p}\right)_m \right]^*
\end{equation*}
where $\left(\dfrac{n}{p}\right)_m$ is defined by
\begin{equation*}
\left(\frac{n}{p}\right)_4 = \begin{cases}
    1 & \textrm{if $n^{(p-1)/m} \equiv 1 \mod p$} \\
    \zeta & \textrm{if $n^{(p-1)/m} \equiv c \mod p$} \\
    \zeta^2 & \textrm{if $n^{(p-1)/m} \equiv c^2 \mod p$} \\
    \zeta^3 & \textrm{if $n^{(p-1)/m} \equiv c^3 \mod p$} \\
        & \vdots \\
    \zeta^{m-1} & \textrm{if $n^{(p-1)/m} \equiv c^{m-1} \mod p$}
\end{cases}
\end{equation*}
Here, $\zeta$ is a primitive $m$-th root of unity in $\complexes$,
and $c$ is defined as a primitive $m$-th root of unity in the multiplicative group of integers mod $p$,
i.e. an integer $c$ such that $c^m \equiv 1 \mod p$ and $c^k \not\equiv 1 \mod p$ for all positive integers $k < m$.

While this $f$ is not an eigenvector of the DFT, it still satisfies the property that each entry except for the first has magnitude 1
and that the magnitude of each entry remains fixed under the DFT.
In particular, there exists a complex constant $z$ of magnitude 1 such that $Wf = z\overline{f}$.

As an example of this construction, let $w$ and $w^*$ be such that $w^2=i$ and $(w^*)^2=-i$.
Sampling the $17 \times 17$ DFT matrix and employing $U=diag[1,i,-w^*,-1,w^*,w,-w,-i,-i,-w,w,w^*,-1,-w^*,i,1]$
generates an equiangular harmonic frame and a companion equiangular frame for $\complexes^{16}$, where
$\inne{g_j,f_j} = 0$ and $\abs{\inne{g_k, f_l}}^2=\frac{17}{256}$ for $k \neq l$.
\end{rem}

%

\subsection{Security analysis}

Here, we assume that the attacks Eve can carry out against the key distribution are of the type intercept/resend, that is, she measures a fraction of signals sent by Alice and forwards a different state to Bob. In the asymptotic limit of sample size of the qubits transmitted the length R of the key string that can be distilled by Alice and Bob with Eve has zero information is:
\begin {equation}
R = I(A:B) - min\{I(A:E), I(B,E)\}.
\end {equation} 
where the quantity I refers to the mutual information between two parties that quantifies how much knowledge of one party's outcome implies the result of the second party. The best strategy for Eve is to use Alice and Bob's basis $50 \%$ of the time as the expression is symmetric with respect to both of them. Eve can choose only one of the bases that will increase the length of the key by breaking the symmetry. She can use a combination of the strategies to restore the symmetry and at the same time maximize the mutual information with either of the parties. It is desirable to quantify the mutual informations in terms of the quantity q the fraction of the signal that Eve intercepts.

{ Let $\set{f_j}_{j=1}^N$ be an equiangular FUNTF for $\complexes^d$, where $N>d$, of square angle $\alpha = |\langle f_j, f_k\rangle|^2 = \frac{N-d}{d(N-1)}$, $\forall j\ne k$. Suppose $d= 2^n$. Then the space $\complexes^d$ can be described by $n$ qubits. In it, the FUNTF  as defined in the Introduction. 
Let $\set{g_j}_{j=1}^N$ be a companion equiangular frame for $\set{f_j}_{j=1}^N$, so $|\langle g_j , f_k \rangle|^2 = \frac{N}{d(N-1)}(1-\delta_{jk})$.

Alice generates one of the states $f_j$ with equal probabilities, $\frac{1}{N}$, and sends it to Bob. He, in turn, performs a measurement obtaining an outcome $g_k$ ($k\ne j$) with probability $\frac{1}{N-1}$. He publicly announces a set of $N-2$ numbers $l \ne k$. If the set does not contain $j$, then Alice declares success, otherwise the protocol fails. Evidently, it succeeds with probability
\begin{equation}
R_0 = \frac{1}{N-1}~.
\label{eqR0}\end{equation}
When it succeeds, Alice and Bob share the information $(j,k)$ which is an ordered pair. By listening to Bob's announcement, Eve knows the set $\{ j,k\}$, but she does not know the order. Therefore, Alice and Bob have generated one shared secret classical bit which is the order of $j,k$ in the pair $(j,k)$, say
\begin{equation} \epsilon_{jk} = \left\{ \begin{array}{ccc}
0 & , & j>k \\ 1 & , & j<k
\end{array}\right. \end{equation}
To gain advantage, Eve intercepts Alice's signal and performs a measurement. Her outcome agrees with Alice's signal with probability $\frac{d}{N}$. The rest of time, she obtains one of the other $N-1$ states, each with probability $\frac{N-d}{N(N-1)}.$

When Eve and Alice agree, the protocol fails with probability $\frac{N-2}{N-1}$, as in the case of no interference by Eve. When Eve disagrees with Alice, then either one of the two numbers Bob leaves out of his public announcement can match Alice's, so the probability of failure is now $\left( \frac{N-2}{N-1} \right)^2$. Then the probability of Alice announcing success is
\begin{align}
R & = 1- \frac{N-2}{N-1} \frac{d}{N} - \left( \frac{N-2}{N-1} \right)^2 \left( 1 - \frac{d}{N}\right) \notag\\
& = \frac{2N^2 -(d+3)N +2d}{N(N-1)^2}
\end{align}
to be compared with the probability of success \eqref{eqR0} without Eve's interference. The error is
\begin{equation}
\epsilon_R = \frac{R}{R_0} - 1 = \frac{(N-d)(N-2)}{N(N-1)}
\end{equation}
which approaches 100\% as $N$ becomes large. This is only possible in higher-dimensional spaces ($d\gg 1$).

When Eve and Alice disagree, Alice can announce success even though she disagrees with Bob's bit (a fact she is unaware of). This occurs once every $N-1$ times, resulting in an error. Therefore,
\begin{align}
\text{QBER} & = \frac{1}{R} \left( 1 - \frac{d}{N}\right) \frac{1}{N-1}\notag\\
& = \frac{(N-1)(N-d)}{2N^2 -(d+3)N +2d}
\end{align}
Notice that QBER approaches 50\% as $N$ becomes large in higher dimensional spaces.
}

\acknowledgments

G.S.\ acknowledges support from the U.S.\ Office of Naval Research under award number N00014-15-1-2646.

K. A. O.\ was partially supported by a grant from the Simons Foundation $\# 319197$, and the U. S.\ Army Research Office  grant  W911NF1610008.

\bibliography{QKD_Comp_DFT}{}

\end{document}